\def\year{2020}\relax
\documentclass[letterpaper]{article} 

\usepackage[utf8]{inputenc}
\usepackage[T1]{fontenc}
\usepackage{hyphenat}
\usepackage{xspace}
\usepackage{amsmath}
\usepackage{amsfonts}
\usepackage{hyperref}
\usepackage{url}
\usepackage{booktabs}
\usepackage{multirow}
\usepackage{subfigure}
\usepackage{makecell}
\usepackage{caption}
\usepackage{minibox}
\usepackage{bbm}
\usepackage{graphicx}
\usepackage{balance}
\usepackage{mathtools}
\usepackage{color}
\usepackage{marvosym}
\usepackage{ifthen}
\usepackage{textcomp}
\usepackage{enumitem}
\usepackage{verbatim}
\usepackage{algorithm}
\usepackage{algorithmic}
\usepackage{numprint}
\usepackage{balance}

\usepackage{amsthm}
\theoremstyle{plain}

\newcommand{\chatoDisplayMode}[1]{#1}

\definecolor{MyRed}{rgb}{0.6,0.0,0.0} 
\definecolor{MyBlack}{rgb}{0.1,0.1,0.1} 
\newcommand{\inred}[1]{{\color{MyRed}\sf\textbf{\textsc{#1}}}}
\newcommand{\frameit}[2]{
  \begin{center}
  {\color{MyRed}
  \framebox[.9\columnwidth][l]{
    \begin{minipage}{.85\columnwidth}
    \inred{#1}: {\sf\color{MyBlack}#2}
    \end{minipage}
  }\\
  }
  \end{center}
}

\newcommand{\note}[2][]{\chatoDisplayMode{\def\@tmpsig{#1}\frameit{{\Pointinghand} Note}{#2\ifx \@tmpsig \@empty \else \mbox{ --\em #1}\fi}}}
\newcommand{\todo}[2][]{\chatoDisplayMode{\def\@tmpsig{#1}\frameit{{\Writinghand} To-do}{#2\ifx \@tmpsig \@empty \else \mbox{ --\em #1}\fi}}}

\newcommand{\abbrevStyle}[1]{#1}

\newcommand{\ie}{\abbrevStyle{i.e.}\xspace}
\newcommand{\eg}{\abbrevStyle{e.g.}\xspace}

\newcommand{\vs}{\abbrevStyle{vs.}\xspace}

\newcommand{\xhdr}[1]{\vspace{1.7mm}\noindent{{\bf #1.}}}

\newcommand{\textcite}[1]{\citeauthor{#1} \shortcite{#1}}

\newcommand{\hide}[1]{}

\makeatletter
\newcommand{\iffont}[2]{\ifthenelse{\equal{\f@family}{#1}}{#2}{}}
\makeatother

  \usepackage{mathptmx}

  \DeclareSymbolFont{greek}{OML}{cmm}{m}{n}
  \DeclareMathSymbol{\alpha}{\mathalpha}{greek}{"0B}
  \DeclareMathSymbol{\beta}{\mathalpha}{greek}{"0C}
  \DeclareMathSymbol{\gamma}{\mathalpha}{greek}{"0D}
  \DeclareMathSymbol{\delta}{\mathalpha}{greek}{"0E}
  \DeclareMathSymbol{\epsilon}{\mathalpha}{greek}{"0F}
  \DeclareMathSymbol{\zeta}{\mathalpha}{greek}{"10}
  \DeclareMathSymbol{\eta}{\mathalpha}{greek}{"11}
  \DeclareMathSymbol{\theta}{\mathalpha}{greek}{"12}
  \DeclareMathSymbol{\iota}{\mathalpha}{greek}{"13}
  \DeclareMathSymbol{\kappa}{\mathalpha}{greek}{"14}
  \DeclareMathSymbol{\lambda}{\mathalpha}{greek}{"15}
  \DeclareMathSymbol{\mu}{\mathalpha}{greek}{"16}
  \DeclareMathSymbol{\nu}{\mathalpha}{greek}{"17}
  \DeclareMathSymbol{\xi}{\mathalpha}{greek}{"18}
  \DeclareMathSymbol{\pi}{\mathalpha}{greek}{"19}
  \DeclareMathSymbol{\rho}{\mathalpha}{greek}{"1A}
  \DeclareMathSymbol{\sigma}{\mathalpha}{greek}{"1B}
  \DeclareMathSymbol{\tau}{\mathalpha}{greek}{"1C}
  \DeclareMathSymbol{\upsilon}{\mathalpha}{greek}{"1D}
  \DeclareMathSymbol{\phi}{\mathalpha}{greek}{"1E}
  \DeclareMathSymbol{\chi}{\mathalpha}{greek}{"1F}
  \DeclareMathSymbol{\psi}{\mathalpha}{greek}{"20}
  \DeclareMathSymbol{\omega}{\mathalpha}{greek}{"21}
  \DeclareMathSymbol{\varepsilon}{\mathalpha}{greek}{"22}
  \DeclareMathSymbol{\vartheta}{\mathalpha}{greek}{"23}
  \DeclareMathSymbol{\varpi}{\mathalpha}{greek}{"24}
  \DeclareMathSymbol{\varrho}{\mathalpha}{greek}{"25}
  \DeclareMathSymbol{\varsigma}{\mathalpha}{greek}{"26}
  \DeclareMathSymbol{\varphi}{\mathalpha}{greek}{"27}
  \DeclareSymbolFont{otone}{OT1}{cmr}{m}{n}
  \DeclareMathSymbol{\Gamma}{\mathalpha}{otone}{0}
  \DeclareMathSymbol{\Delta}{\mathalpha}{otone}{1}
  \DeclareMathSymbol{\Theta}{\mathalpha}{otone}{2}
  \DeclareMathSymbol{\Lambda}{\mathalpha}{otone}{3}
  \DeclareMathSymbol{\Xi}{\mathalpha}{otone}{4}
  \DeclareMathSymbol{\Pi}{\mathalpha}{otone}{5}
  \DeclareMathSymbol{\Sigma}{\mathalpha}{otone}{6}
  \DeclareMathSymbol{\Upsilon}{\mathalpha}{otone}{7}
  \DeclareMathSymbol{\Phi}{\mathalpha}{otone}{8}
  \DeclareMathSymbol{\Psi}{\mathalpha}{otone}{9}
  \DeclareMathSymbol{\Omega}{\mathalpha}{otone}{10}
  \DeclareSymbolFont{syms}{OML}{cmm}{m}{it}
  \DeclareMathSymbol{\partial}{\mathord}{syms}{"40}
  \DeclareMathAlphabet{\mathbold}{OML}{cmm}{b}{it}
  \DeclareSymbolFont{largesymbols}{OMX}{cmex}{m}{n}

\usepackage{aaai20}  
\usepackage{times}  
\usepackage{helvet} 
\usepackage{courier}  
\usepackage{graphicx} 
\usepackage{graphicx}  
\nocopyright 

\title{Adoption of Twitter's New Length Limit: Is 280 the New 140?}

\author{
  Kristina Gligori\'c\\
  EPFL\\
  kristina.gligoric@epf\/l.ch
  \And
  Ashton Anderson\\
  University of Toronto\\
  ashton@cs.toronto.edu
  \And
  Robert West\\
  EPFL\\
  robert.west@epf\/l.ch
}

\begin{document}

\maketitle

\begin{abstract}
 
In November 2017, Twitter doubled the maximum allowed tweet length from 140 to 280 characters, a drastic switch on one of the world's most influential social media platforms.
In the first long-term study of how the new length limit was adopted by Twitter users, we ask:
Does the effect of the new length limit resemble that of the old one?
Or did the doubling of the limit fundamentally change how Twitter is shaped by the limited length of posted content?
By analyzing Twitter's publicly available 1\% sample over a period of around 3 years, 
we find that, when the length limit was raised from 140 to 280 characters, the prevalence of tweets around 140 characters dropped immediately, while the prevalence of tweets around 280 characters rose steadily for about 6 months.
Despite this rise, tweets approaching the length limit have been far less frequent after than before the switch.
We find widely different adoption rates across languages and client-device types.
The prevalence of tweets around 140 characters before the switch in a given language is strongly correlated with the prevalence of tweets around 280 characters after the switch in the same language, and very long tweets are vastly more popular on Web clients than on mobile clients.
Moreover, tweets of around 280 characters after the switch are syntactically and semantically similar to tweets of around 140 characters before the switch, manifesting patterns of message squeezing in both cases.
Taken together, these findings suggest that the new 280\hyp character limit constitutes a new, less intrusive version of the old 140\hyp character limit.
The length limit remains an important factor that should be considered in all studies using Twitter data.

\end{abstract}

\section{Introduction}
\label{sec:intro}
\noindent 
On 7 November 2017, Twitter suddenly and unexpectedly increased the maximum allowed tweet length from 140 to 280 characters, thus altering its signature feature.
According to Twitter, this change, which we henceforth refer to as ``the switch'', was introduced to give users more space to express their thoughts, as a disproportionately large fraction of tweets had been exactly 140 characters long \cite{t2,w1}.
Understanding the consequences of the switch is of paramount importance for social media studies, for two reasons:
first, because Twitter is one of the leading social media platforms \cite{perrin2019share}, with content posted there reaching and affecting billions of people across the globe;
and second, because the constraints that a medium imposes affect the audience not only through the content delivered over the medium, but also through the characteristics of the medium itself, or, in a mantra coined by Marshall McLuhan \shortcite{mcluhan1964extensions}, ``the medium is the message''.
Thus, anybody in whose research Twitter plays a role cannot ignore the switch, and must consider how the new length limit has impacted Twitter as a platform.

Early work that studied Twitter users' attitudes toward the new 280-character limit \cite{10.1145/3293339.3293349} discovered varying initial reactions ranging from anticipation, surprise, and joy to anger, disappointment, and sadness. 
Early studies also revealed a low initial prevalence of long tweets immediately following the switch \cite{twitter280blog}
and studied the short-term impact that the length limit had on linguistic features and engagement \cite{w1,boot2019character}, also in the specific context of political tweets \cite{jaidka2019brevity}.

Whereas the above-cited, early studies necessarily had to consider short-term effects, much less is known about the long-term effects of the switch.
Now, nearly three years after the switch, the present paper constitutes the first attempt to bridge this gap with a long-term study spanning several years.
Broadly, our research was guided by the question, \textit{Is 280 the new 140?}
In other words, does the effect of the new length limit resemble that of the old one, just over a broader range of potential tweet lengths?
Or has the doubling of the limit fundamentally changed how Twitter is shaped by the limited length of posted content?

\xhdr{Research questions}
Concretely, we address the following research questions:

\textbf{RQ1:} How did tweet length change in the two years following the switch?

\textbf{RQ2:} How did tweet length change across languages?

\textbf{RQ3:} How did tweet length change across devices (Web \vs\ mobile clients \vs automated sources)?

\textbf{RQ4:} What are the syntactic and semantic characteristics of long \vs\ short tweets?

\begin{figure}
    \centering
    \includegraphics[width= 0.34\textwidth]{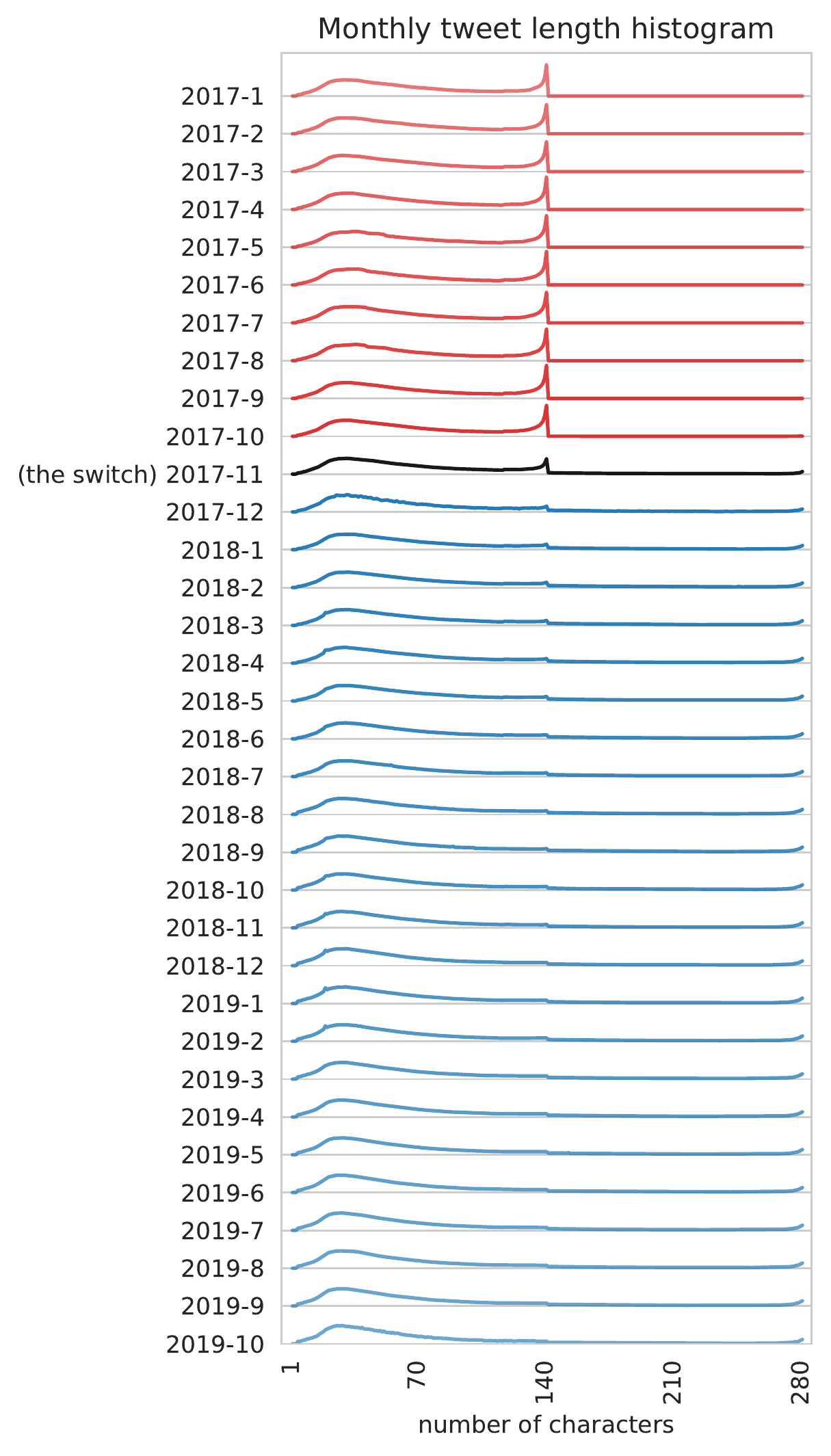}
    \caption{Monthly tweet length histograms, normalized.}
    \label{fig1}
\end{figure}

\xhdr{Summary of main findings}
Using a 1\% sample of all tweets spanning the period from 1 January 2017 to 31 October 2019,
we conduct an observational study of tweet length over time (RQ1).
Once the length limit was doubled and the old 140-character limit became obsolete, we find a decline in the prevalence of tweets of exactly or just under 140 characters, and a rise in the prevalence of tweets of exactly or just under 280 characters, with a smooth adaptation phase of around 6 months.

Comparing languages (RQ2), we find vastly different levels of adoption patterns. The prevalence of 140 characters before the switch is strongly correlated with the prevalence of 280 characters after the switch, indicating that some languages have an inherent affinity to longer messages.
Comparing device types (RQ3), tweets above 140 characters are used more on Web clients, compared to mobile clients. Automated sources and third-party applications were the slowest to adapt. They continue to tweet around 140 characters disproportionately more often, compared to regular users, and in general, tend to publish longer tweets.

Finally, we observe that 280\hyp character tweets are syntactically and semantically similar to 140\hyp character tweets posted before the switch (RQ4). The 280\hyp character tweets show linguistic fingerprints indicative of ``message squeezing'', with inessential parts of speech (\eg, fillers, adverbs, conjunctions) being relatively less frequent, and essential parts of speech (\eg, verbs, negations) being relatively more frequent, compared to shorter tweets. Their usage is additionally associated with specific topics, such as Money, Death, Work, and Religion.

In a nutshell, although the doubling of the length limit eliminated the drastic disproportion of tweets reaching the maximum length (\eg, 9\% of English tweets used to be exactly 140 characters long before the switch), our results demonstrate the emergence of a similar, though considerably weaker, effect around 280 characters after the switch.
We hence answer the guiding question---\textit{Is 280 the new 140?}---in a nuanced way:
\textit{280 is a less intrusive 140.}
These findings have important implications for Twitter\hyp based research, as they show that, although the new limit is ``felt less'' by users than the old limit, 280 characters still constitutes an impactful length constraint that shapes the nature of Twitter.

\section{Related work}
\label{sec:related}

\begin{figure*}[ht]
\begin{minipage}[c][13cm][t]{.23\textwidth}
  \centering
  \includegraphics[height=9.5cm]{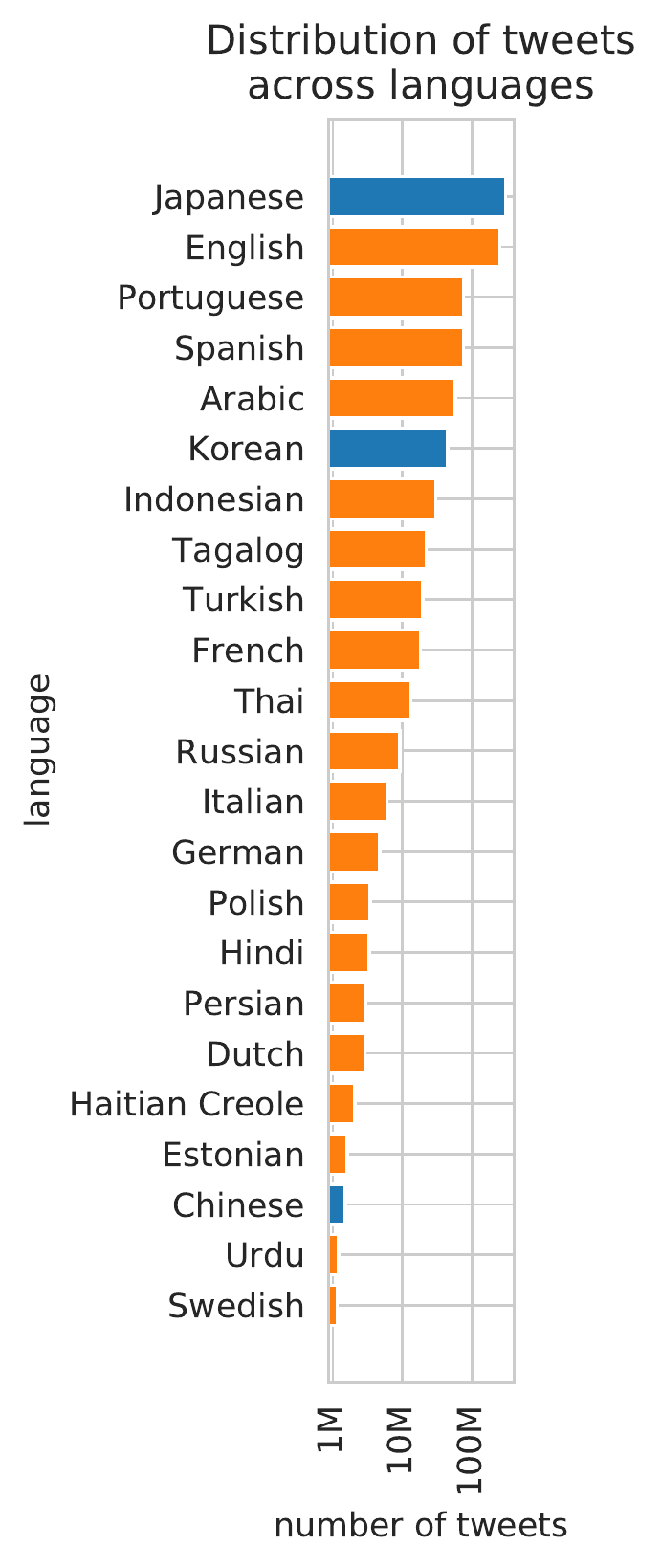}
  \caption{Number of original tweets in the 1\% sample posted between January 1st 2019 and October 31st 2019, across the 23 studied languages where the switch happened (in orange), and did not happen (in blue)}
  \label{fig:languages}
\end{minipage}%
\hspace*{0.7cm}
\begin{minipage}[c][13cm][t]{.73\textwidth}
  \centering
  \includegraphics[height=5cm]{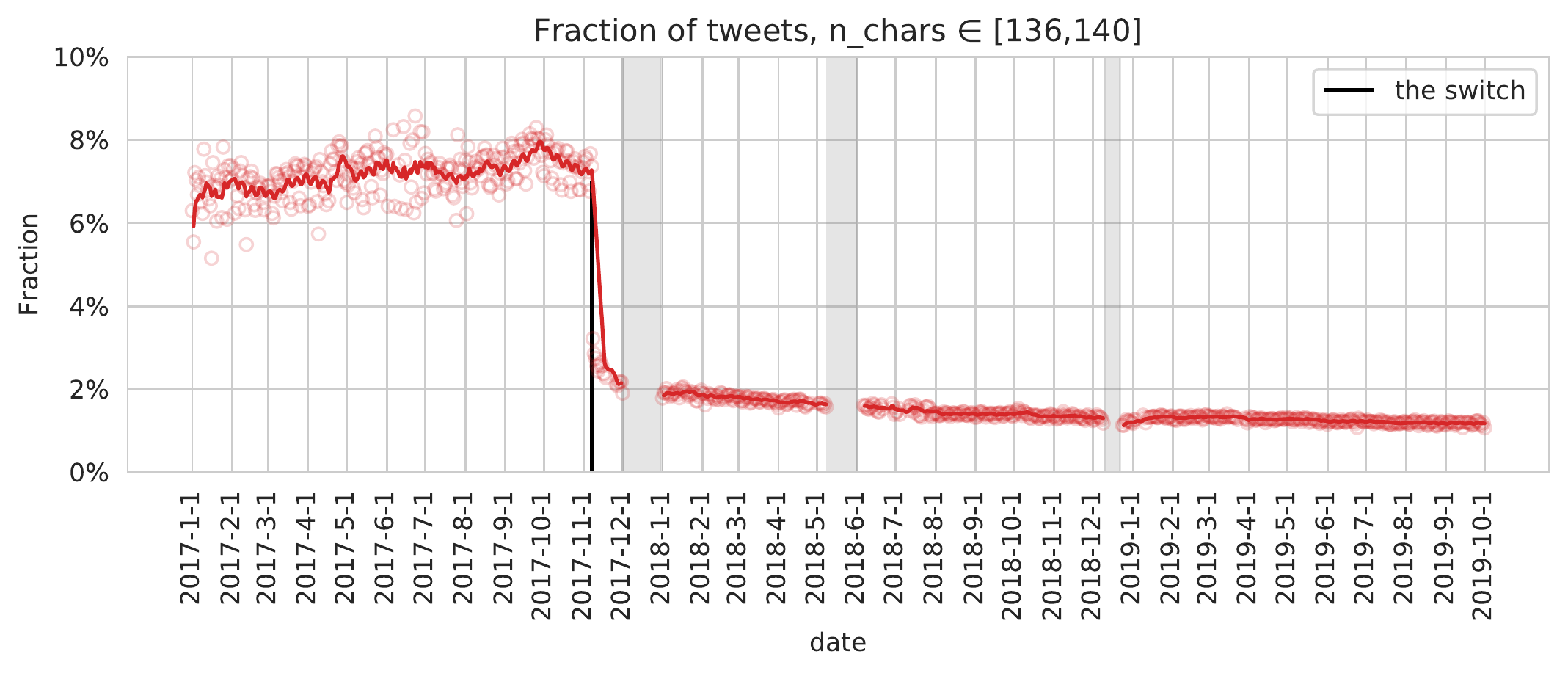}
  \caption{Daily fraction (indicated with a circle), and 10-day rolling average (solid line) of faction of tweets that have between $136$ and $140$ characters (inclusive).}
  \label{fig2}\par\vspace{0.25cm}
  \includegraphics[height=5cm]{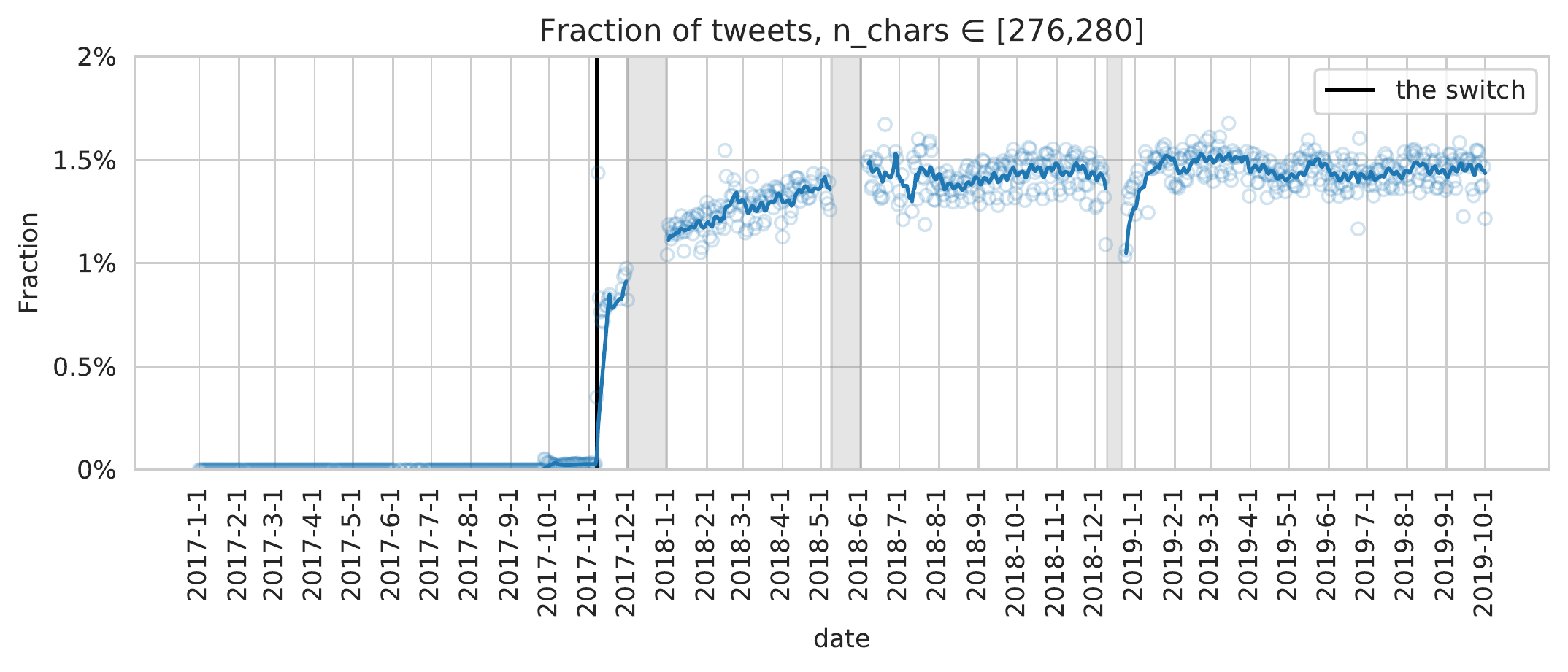}
  \caption{Daily fraction (indicated with a circle), and 10-day rolling average (solid line) of faction of tweets that have between $276$ and $280$ characters (inclusive). Note the different y-axis scales.}
  \label{fig3}
\end{minipage}

\end{figure*}

\xhdr{Twitter communication and supporting features: studies of use and emerging conventions} Previous work has extensively studied communication taking place on Twitter and the specific features that support them, most importantly: retweets \cite{boyd2010tweet}, hashtags \cite{wikstrom2014srynotfunny,page2012linguistics}, quotes \cite{garimella2016quote}, and emojis \cite{pavalanathan2016more}. Previous work has also investigated linguistic conventions on Twitter, the patterns of their emergence \cite{kooti2012emergence,kooti2012predicting}, how users align to them in conversations \cite{doyle2016robust}, how they diffuse \cite{centola2018experimental,chang2010new}, and how they continuously evolve \cite{cunha2011analyzing}.

\xhdr{Variations in usage and adoption of conventions and linguistic style} Additionally, previous work has studied how patterns of adoption of these features, as well as the linguistic style used on the platform more broadly, varies across numerous dimensions \cite{shapp2014variation}, including gender \cite{ciot2013gender}, political leaning \cite{sylwester2015twitter}, age and income \cite{flekova2016exploring}; but also within accounts 
\cite{clarke2019stylistic}.

\xhdr{Length limit and the impact of message length on success and linguistic characteristics} Previous work has studied how the imposed length constraint on Twitter and other microblogging platforms affects the dialogues and the linguistic style \cite{zhou2019remaking,jin2017will}, and the success of the message measures through the received engagement \cite{r9,w1,10.1145/3359147,wang2020does,wasike2013framing}. More broadly, philology, communication, education, and psychology scholars have investigated conciseness and its benefits in many different contexts \cite{laib1990conciseness,vardi2000brevity,sloane2003say}.

\xhdr{Message framing on Twitter} Many previous studies have investigated the question of what wording makes messages successful in online social media, often formulated as the task of predicting what makes textual content become popular \cite{berger2012makes,guerini2011exploring,lamprinidis2018predicting}. In the specific case of Twitter, in addition to characterizing how language is used on the platform in general \cite{murthy2012towards,levinson2011long,hu2013dude,eisenstein2013bad}, researchers have investigated the correlation of linguistic signals with the propagation of tweets \cite{artzi2012predicting,bakshy2011everyone,r9,doi:10.1080/15534510.2016.1265582}.



\section{Data}
\label{sec:matmet}
We use publicly available 1\% sample of tweets, spanning the period between 1 January 2017, and 31 October 2019, available on the Internet Archive.%
\footnote{\url{https://archive.org/details/twitterstream}}
We consider original tweets (\ie, we discard retweets). There are between 1 and 1.5 million daily original tweets. In Figure~\ref{fig:languages} we show the exact number in total across languages. 

We study 23 biggest languages: three languages where the switch did not happen: Japanese, Korean, and Chinese, and twenty where it happened, each language with more than 2M tweets in total. The switch did not happen in Japanese, Korean, and Chinese because the 140 characters limit was not as restrictive as it was in other languages, since more information can be conveyed with the same number of characters \cite{twitter_blog1,twitter_blog2}. We note that the data is sampled at the community level. We stay at describing the community level change as opposed to user-level behaviors since user-level information is incomplete (in expectation, we have 1\% of tweets posted by a fixed user). 

\xhdr{Character counting} We carefully count the number of characters based on the official documentation%
\footnote{\url{https://developer.twitter.com/en/docs/counting-characters}}. Tweet length is counted using the Unicode normalization of the tweet text. The tweet text is selected from the tweet object using \textit{displayed text range} information, discarding any retweet tags, and leading @user mentions that are not counted towards the length limit. The text content of a Tweet can contain up to 140 characters (or Unicode glyphs) before the switch, or 280 after the switch. Emoji sequence using multiple combining glyphs counts as multiple characters. Glyphs used in Chinese, Japanese or Korean languages are counted as one character before, and as two characters after the introduction of the 280 limit. Therefore, a Tweet composed of only CJK text can only have a maximum of 140 of these types of glyphs after the switch.

\begin{figure*}
    \begin{minipage}{\textwidth}
        \centering
        \includegraphics[width=0.99\textwidth]{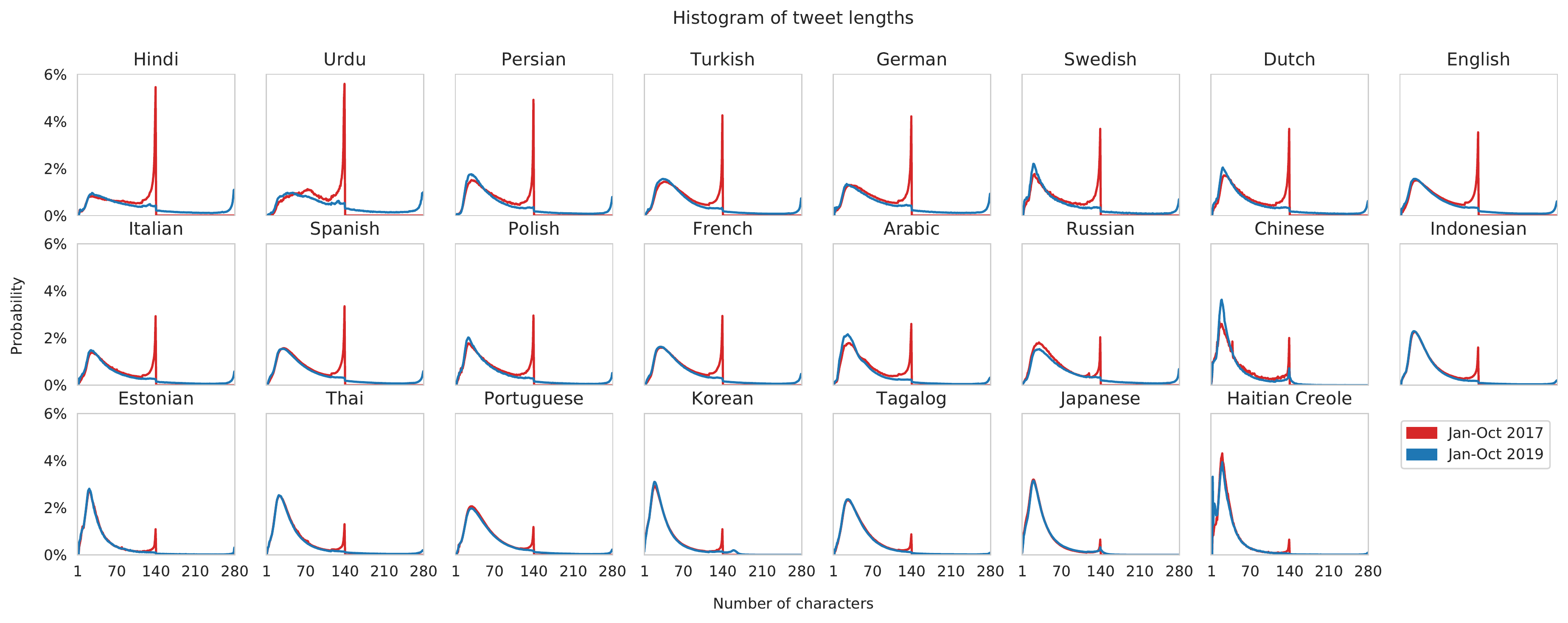}
    \end{minipage}

    \caption{Tweet length histograms in the $23$ studied languages, for the period before the 280 limit was introduced (in red), and after it was introduced (in blue). The languages are sorted by the prevalence of 140 before the switch.}
    \vspace{-2mm}
    \label{fig4}
\end{figure*}

\begin{figure}
    \centering
    \includegraphics[width= 0.43\textwidth]{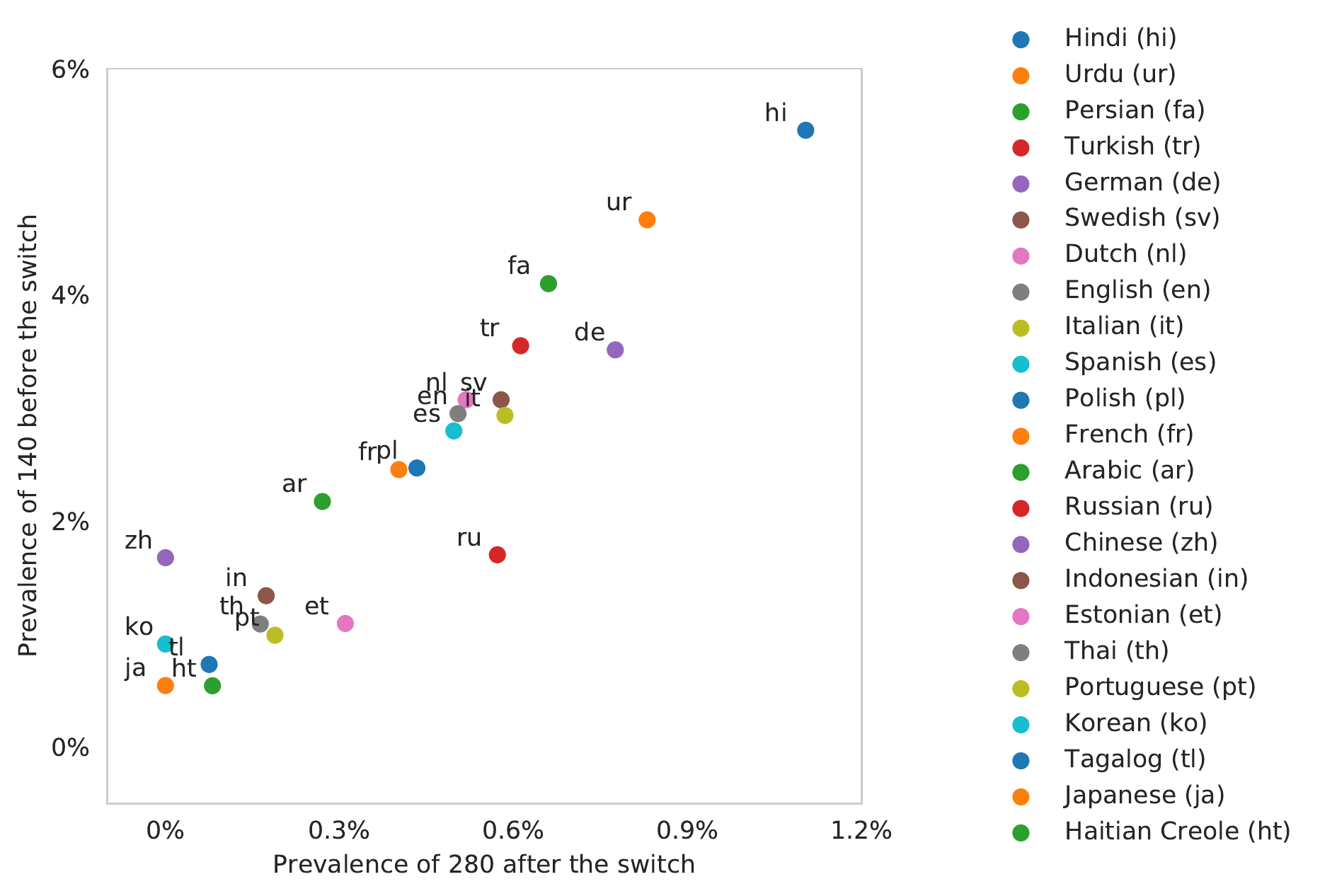}
    \caption{Fraction of tweets that are exactly 280 characters long after the switch, on x-axis, and fraction of tweets that are exactly 140 characters long before the switch, on y-axis (Spearman rank correlation $0.92$, $p = 4.87*10^{-10}$).}
    \label{fig5}
\end{figure}

\section{Results}
\label{sec:results}
\begin{figure*}
    \centering
    \includegraphics[width= 0.99\textwidth]{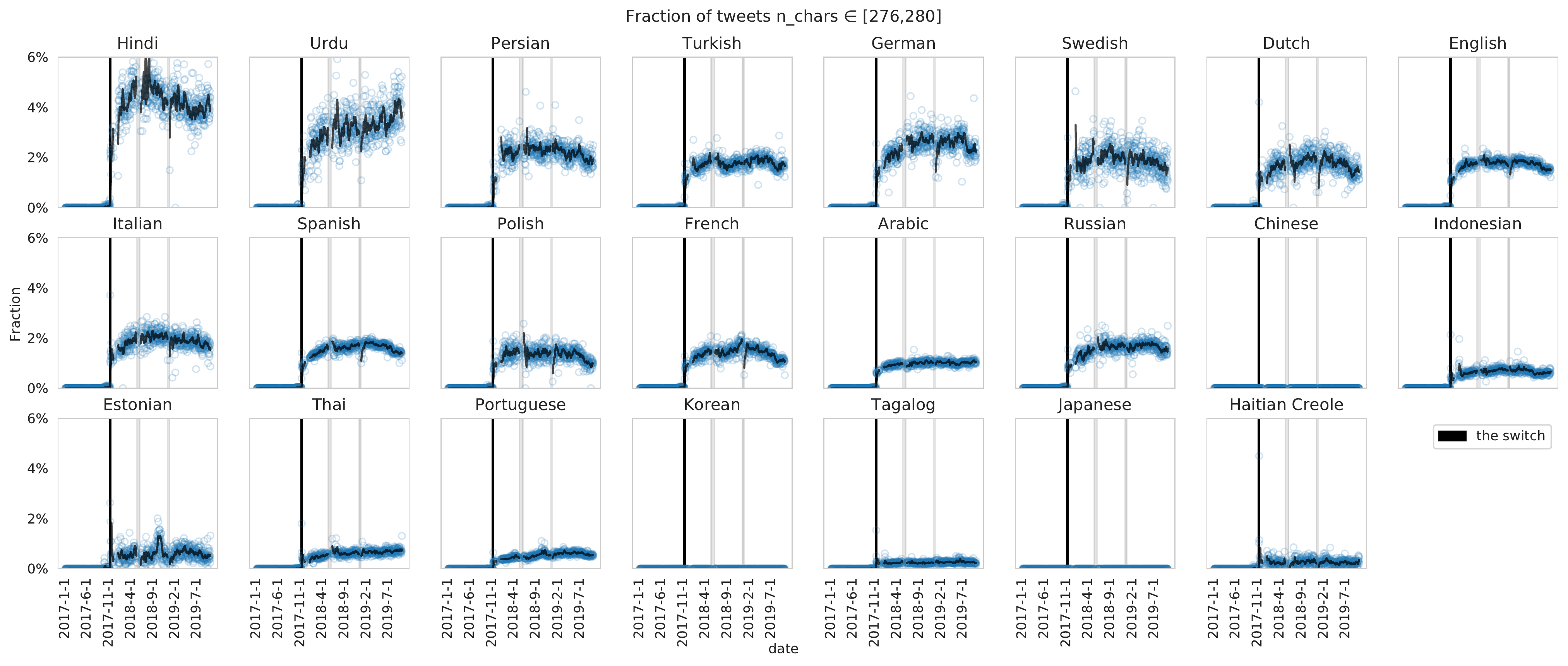}
    \caption{Daily evolution of 280 limit adoption across languages. Daily fraction (indicated with a circle), and 10-day rolling average (solid line) of faction of tweets that have between $276$ and $280$ characters (inclusive). The languages are sorted by prevalence of 140 before the switch.}
    \label{fig7}
\end{figure*}

\subsection{RQ1: How did tweet length change in the two years following the switch?}

First, we measure the prevalence of different tweet lengths over time. We start by inspecting monthly histograms of tweet lengths, shown in  Figure~\ref{fig1}, where across 20 studied languages we visualize the distribution in red for the months before the switch, and in blue for the months after.

We observe a sharp decline of 140-character tweets and an increase in 280-character tweets after the switch. Otherwise, the first peak, consistently between 25 and 30 characters, remains unchanged (\ie, the mode of the distribution is stable).

Next, we focus on these two interesting ranges, close to 140 and close to 280, and do a more granular daily analysis. We monitor the daily fraction of tweets that either hit the character limit or are within a five-character margin of it. The daily fraction of tweets that have between 136 and 140 characters are shown in Figure~\ref{fig2}, and the daily fraction of tweets between 276 and 280 characters in Figure~\ref{fig3}. Here we consider again original tweets across the 20 studied languages, where the 280 character limit was introduced. Gray stripes indicate omitted days with missing data. 

We note that while 140-character tweets became less prevalent after the switch, dropping from 7.0\% to 1.2\% over the studied period, the prevalence of 280-character tweets steadily increased for 6 months after its introduction, reaching 1.5\% at the end of the studied period.
 
\subsection{RQ2: How did tweet length change across languages?}

Next, we seek to characterize the adoption across the studied languages. We start by examining in Figure~\ref{fig4} histograms of tweet lengths in red Jan-Oct 2017 as pre-switch period, and in blue the same months 2 years later when the things settled in Jan-Oct 2019, as post-switch period. Here we restrict ourselves to tweets posted from regular sources (web and mobile interface, rather than third-party applications and automated sources). Similar to the overall view, across languages the first peak and the mode of the distribution are constant, and the interesting character length ranges are near 140 and near 280 characters.

In Figure~\ref{fig5} we show the fraction of tweets that are exactly 280 characters long after the switch, on the x-axis, and the fraction of tweets that are exactly 140 characters long before the switch, on the y-axis.

We observe that the prevalence of 140 before the switch in a language is correlated with the prevalence of 280 after. The more 140 was used in a language before the switch, the more 280 is used after (Spearman rank correlation $0.92$, $p = 4.87*10^{-10}$). In Hindi and Urdu, 280 is very prevalent, and it is the mode of the distribution--the most frequent character length after the switch.

In Figure~\ref{fig7}, we further monitor the evolution of adoption patterns of the new limit across languages, for tweets posted from web and mobile sources. In most of the languages, the prevalence seems to have settled, and is even decreasing again, \ie, the peak of usage is past. Urdu is a notable exception, where the prevalence is still growing, and the adoption rate is still not in a stable state.

\xhdr{Differences in differences estimation of the effect of switch}
Additionally, we take advantage of the fact that the new limit was not introduced in all languages to perform a differences in differences estimation of the effect of the switch on tweet lengths.

\begin{figure}[t]
    \centering
    \includegraphics[width= 0.35\textwidth]{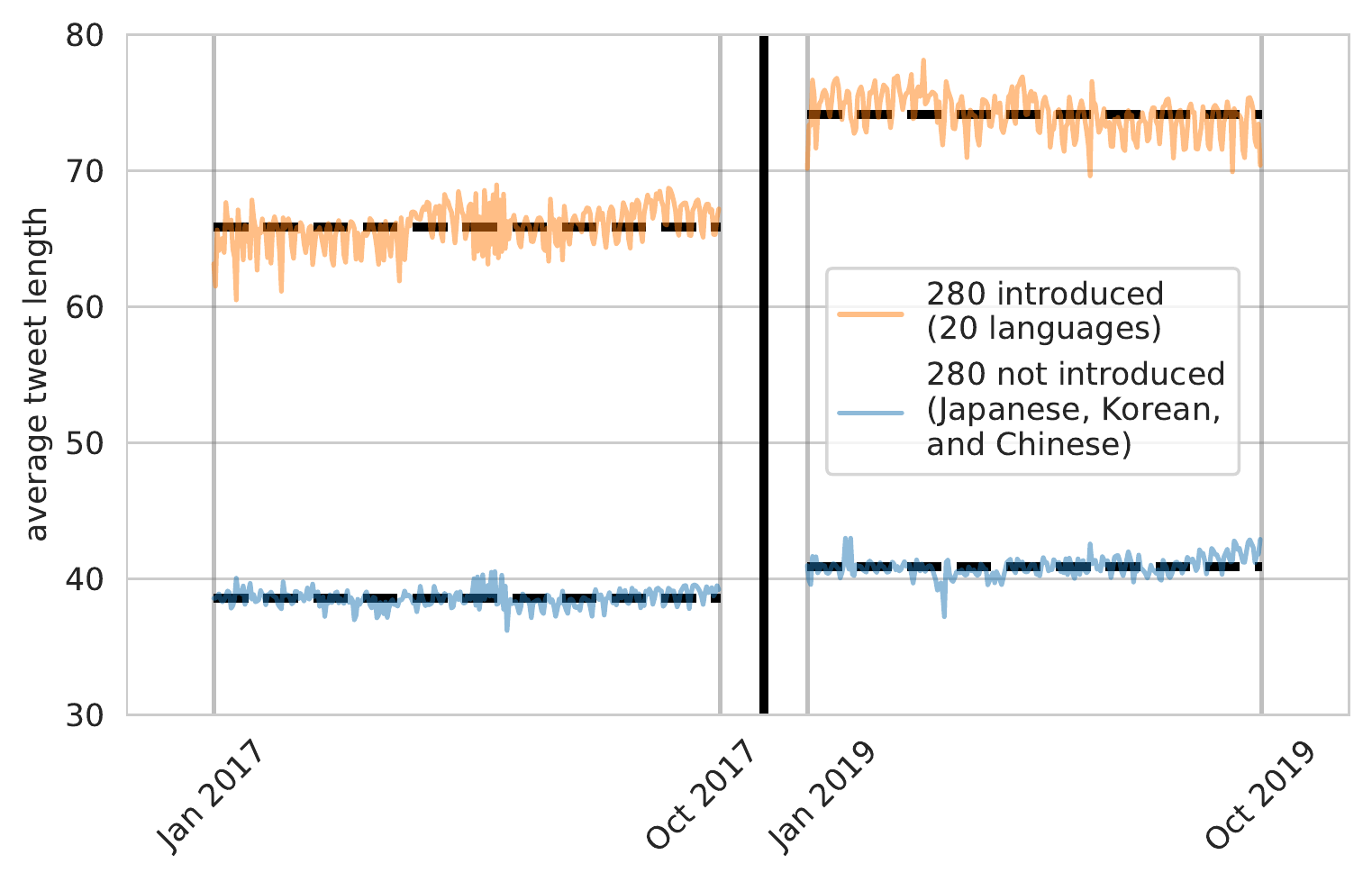}
    \caption{Differences in differences setup for estimation of the effect of the introduction of 280 character limit on tweet lengths. For the pre- (left) and post-switch (right) periods, we monitor the daily average tweet length, in the languages where the 280 limit is introduced (in orange), and in the language where it is not introduced (in blue). Dashed lines mark the averages of the daily average tweet length in the four conditions.}
    \label{fig9}
\end{figure}

To go beyond visual inspection and to account for possible global platform-wide changes that are not associated with the switch, we use a differences in differences regression estimation, where the tweet lengths in Japanese, Korean and Chinese, languages where the 280 characters were not introduced are the control timeseries, and the tweet lengths in the other 20 studied languages (Figure ~\ref{fig:languages}) are the treated timeseries. Both are observed in the pre-switch (Jan-Oct 2017), and post-switch (Jan-Oct 2019) periods, as illustrated in Figure~\ref{fig9}.
We fit a model
\begin{equation}
    y \sim \mathtt{treated * period},
    \label{eqn:formula_overall}
\end{equation}
where the dependent variable $y$ is the logarithm of the average tweet length for each studied calendar day,
and as independent variables are the following two factors:
$\mathtt{treated}$ (indicates whether the switch was introduced or not in those languages),
$\mathtt{period}$ (indicates whether a calendar day is in year pre-switch or post-switch). 
$\mathtt{treated * period}$ is shorthand notation for $\alpha + \beta \mathtt{treated} + \gamma \mathtt{period} + \delta \mathtt{treated:period} + \epsilon$, where in turn $\mathtt{treated:period}$ stands for the interaction of $\mathtt{treated}$ and $\mathtt{period}$.

The interaction term $\mathtt{treated:period}$ $\delta$ is then the effect of switch on the logarithm of average tweet length. Each studied pre- or post-switch period spans 277 days per condition, amounting to a total of $4 \times 277 =$ 1108 datapoints. The model is multiplicative due to the log. The relative increase over the baseline is then calculated by converting back to the linear scale the fitted coefficient $\delta$. Fitting the model~\ref{eqn:formula_overall}, we measure a $e^{\delta}-1 = e^{0.0598}-1 = 6.16\%$ (95\% CI [5.68\%, 6.64\%]) increase in tweet lengths in the languages where the switch happened, over the control baseline.

To conclude, the estimate a significant increase in tweet lengths in languages where the switch happened, compared to the control languages.

\subsection{RQ3: How did tweet length change across devices?}

\begin{figure}[t]
    \centering
    \includegraphics[width= 0.47\textwidth]{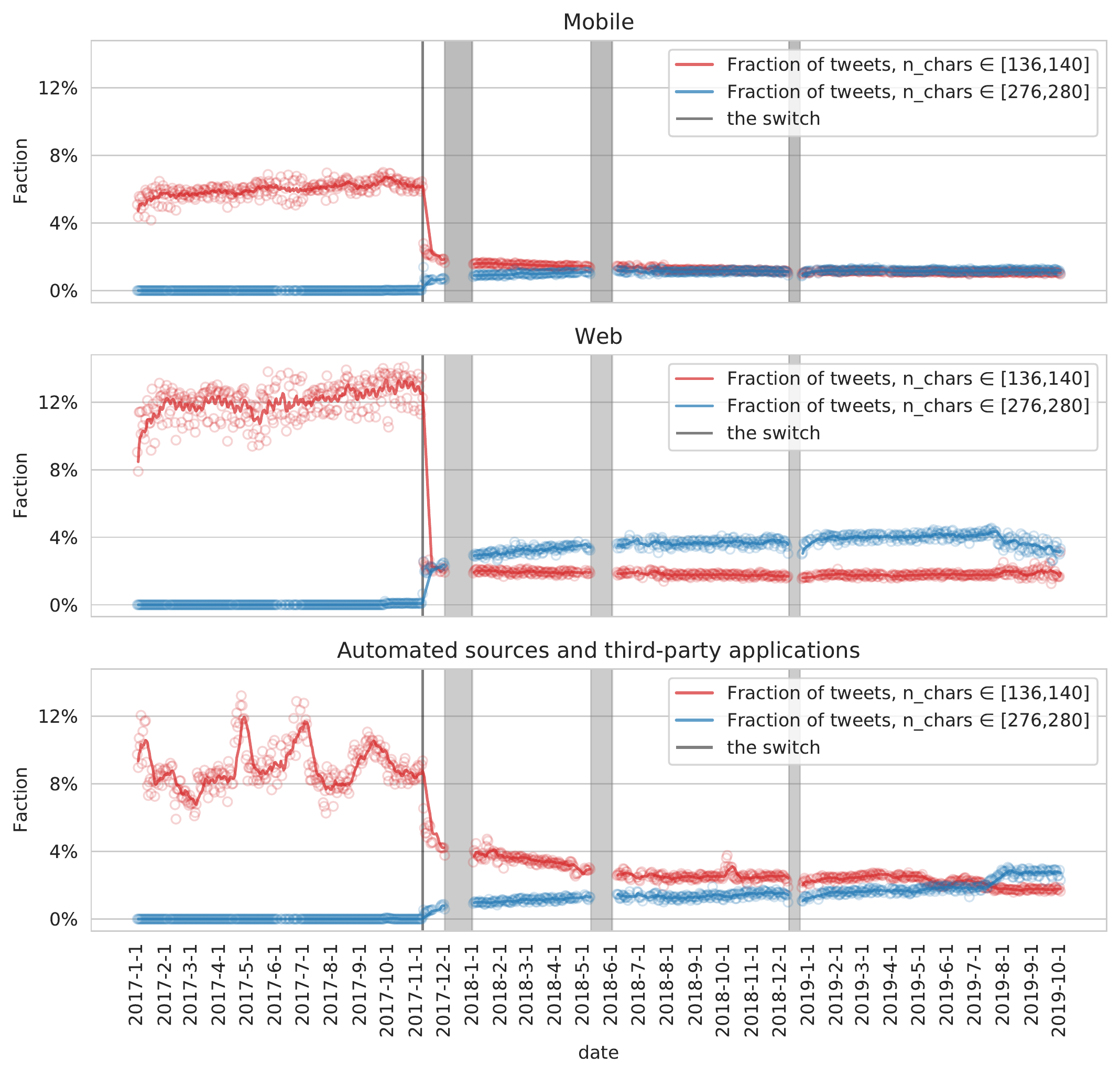}
    \caption{ Daily fraction (indicated with a circle), and 10-day rolling average (solid line) of faction of tweets that have between 276 and 280 characters (blue), and between 136 and 140 characters (red). The evolution is shown separately for Mobile applications, Web interface, and Automated sources and third-party applications.}
    \label{fig8}
\end{figure}

To understand the adoption of the new limit across different devices, we monitor the evolution of the daily fraction of tweets in interesting tweet lengths separately across web, mobile, and automated sources and third-party applications, in Figure~\ref{fig8}. The tweets are tweeted in the 20 languages where the switch happened.

We observe different adoption patterns between web and mobile. Longer tweets are used more on the web. In the web interface, where 140 was most prevalent (around 12\%), 140 was quickly surpassed by 280, reaching around 4\% at the end of the studied period. Automated sources and third-party applications the slowest to adapt.

Tweet length of 280 characters and the near long lengths surpassed 140 around June 2018 (~7 months after the switch) for mobile, but for web immediately after the switch.

\xhdr{Differences in differences estimation of the effect of switch on tweets posted from different devices}
To understand the impact of the switch on tweets posted from different devices, we fit a slightly different model
\begin{equation}
    y \sim \mathtt{treated * period * source},
    \label{eqn:formula_sources}
\end{equation}
where the $\mathtt{source}$ is a categorical variable representing mobile devices, web devices, or automated sources and third-party applications. By analogy to Equation~\ref{eqn:formula_overall}, we then isolate the total effect of switch on tweet length posted by specific sources as the sum of the baseline effect of the switch and the source-specific switch effect,  $\mathtt{treated:period + treated:period:source}$. Fitting the model from Equation~\ref{eqn:formula_sources}, we consistently measure a largest increase in tweet length of 17.46\% (95\% CI [16.54\%, 18.39\%]) for web, followed by 9.76\% (95\% CI [8.88\%, 10.66\%]) for automated sources and third-party applications, and the smallest of 5.64\% (95\% CI [5.19\%, 6.08\%]) for mobile.

\begin{figure}[t]
    \centering
    \includegraphics[width= 0.49\textwidth]{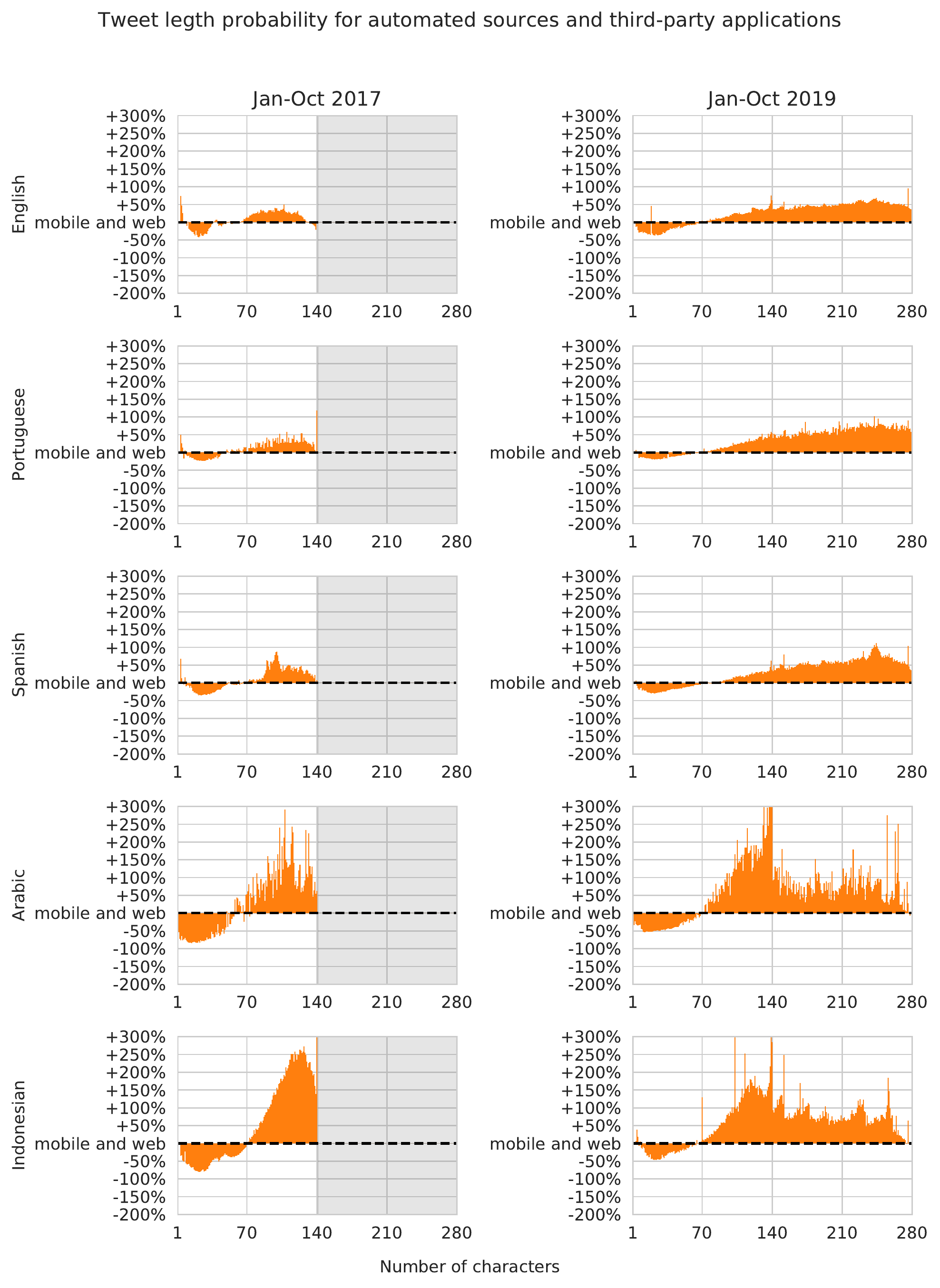}
    \caption{Tweet length probability for tweets posted by automated sources and third-party applications, before the 280 limit was introduced (on the left), and after (on the right). The probability is normalized relative to the baseline, the probability of same length in the same language for regular sources: web and mobile.}
    \label{fig6}
\end{figure}

Next, we further investigate tweets posted by automated sources and third-party applications. This is content likely generated by bots or otherwise automated applications, as opposed to the content generated by regular users. We show for five biggest languages where 280 was introduced (English, Portuguese, Spanish, Arabic, and Indonesian), in Figure~\ref{fig6}, tweet length histograms of tweets posted by third-party applications and automated sources. The probability is normalized relative to the baseline, the probability of the same character length in the same language for regular sources (web and mobile). For example, +200\% means that a tweet of the given length is two times more likely to be observed among the tweets posted by automated sources, compared to the regular users. Again, we compare Jan-Oct 2017 and 2019 as pre- and post-periods. The patterns look similar across languages. Automated sources and third-party applications tweet longer tweets before the switch, and post tweets in longer length ranges after the switch. The infliction point at around 70 characters remains. Automated sources and third-party applications have a lingering peak at 140 after the switch in English, Arabic, and Indonesian.

\subsection{RQ4: What are the syntactic and semantic characteristics of long \vs\ short tweets?}

Lastly, we aim to provide deeper insights into the nature of long tweets tweeted after the switch. Why do users tweet long tweets? What are their signature characteristics? To answer those questions, we examine the content of the tweets.

We study tweets in English that were posted from mobile and web devices during the two previously introduced pre- (75.56M tweets) and post-switch (65.29M tweets) periods. We annotated the tweets with LIWC \cite{pennebaker2007liwc2007} syntactic features (linguistic categories) and semantic features (psychological, biological, and social categories).

\begin{figure}[t]
    \begin{minipage}{.47\textwidth}
        \centering
        \includegraphics[width=\textwidth]{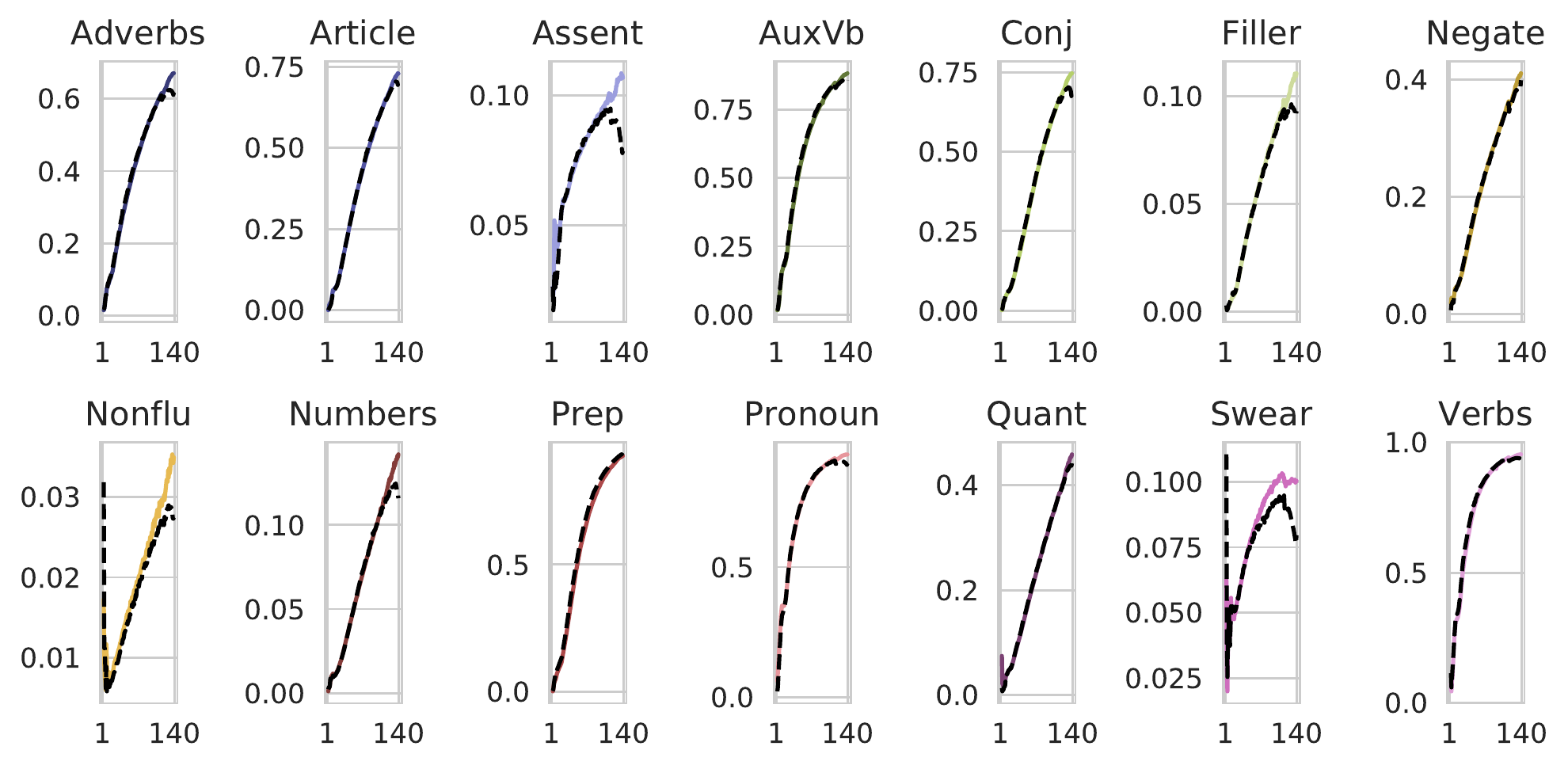}
        \caption{Occurrence frequency of POS tags across tweets in different character lengths possible in the period before switch ($[1-140]$ characters). The dashed black line represents this quantity across tweets posted in the period before the switch (under 140 character limit), and the solid colored line in the period after switch (under 280 character limit).}
        \label{fig:10}
    \end{minipage}
    \begin{minipage}{.47\textwidth}
        \centering
        \includegraphics[width=\textwidth]{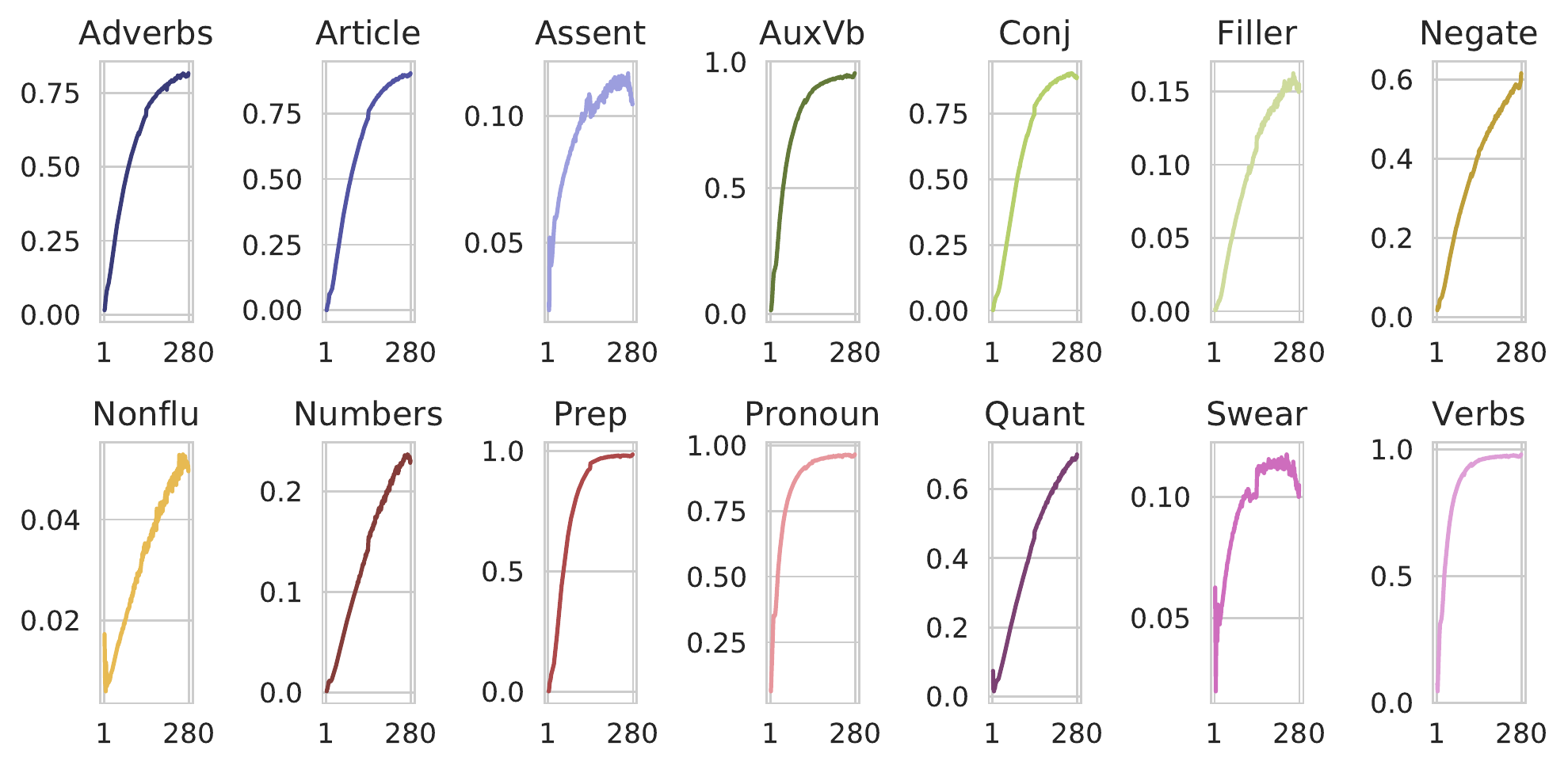}
        \caption{Occurrence frequency of POS tags across tweets posted in the period after the switch (under 280 character limit) in different allowed character lengths ($[1-280]$ characters). POS tags are sorted alphabetically. Note the different y-axis scales.}
        \label{fig:11}
    \end{minipage}
\end{figure}

\begin{figure}[t]
    \begin{minipage}{.47\textwidth}
        \centering
        \includegraphics[width=0.91\textwidth]{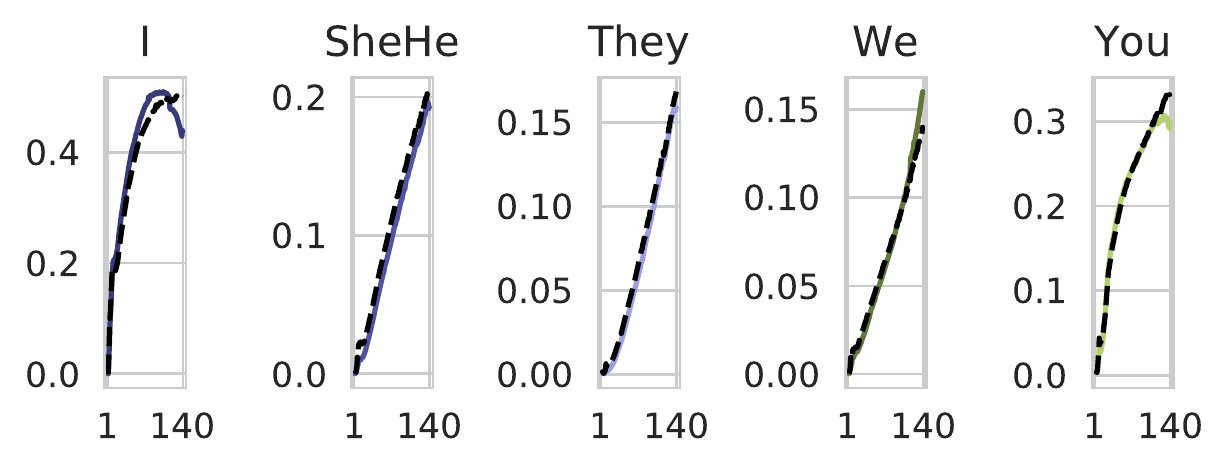}
        
        \caption{Occurrence frequency of personal pronouns across tweets in different character lengths possible in the period before switch ($[1-140]$ characters). The dashed black line represents this quantity across tweets posted in the period before the switch (under 140 character limit), and the solid colored line in the period after switch (under 280 character limit).}
        \label{fig:12}
    \end{minipage}
    \hfill
    \begin{minipage}{.47\textwidth}
        \centering
        \includegraphics[width=0.88\textwidth]{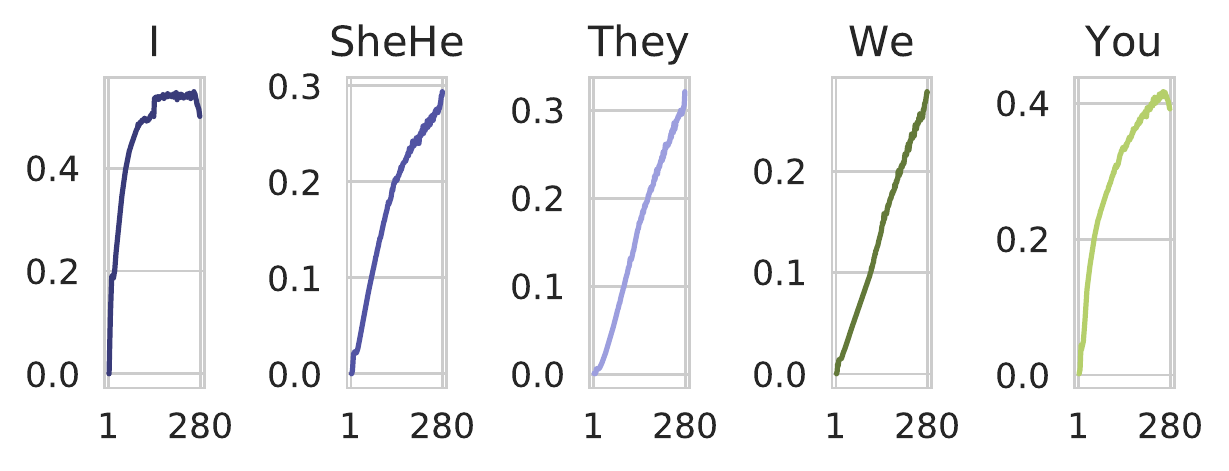}
        \caption{Occurrence frequency of personal pronouns across tweets posted in the period after the switch (under 280 character limit) in different allowed character lengths ($[1-280]$ characters). Personal pronouns are sorted alphabetically. Note the different y-axis scales.}
        \label{fig:13}
    \end{minipage}
\end{figure}

First, to characterize syntactic features of tweets, for all tweets with a given number of characters, we measure the occurrence frequency of different syntactic features--part of speech (POS) tags among tweets of that length. In Figure~\ref{fig:10}, across all possible tweet lengths in the period before the switch ($[1-140]$ characters), we observe the fraction of tweets in that length that have at least one POS tag. The dashed black line represents this quantity across tweets posted in the period before the switch (under 140 character limit), and the solid colored line represents this quantity among tweets posted in the period after the switch (under 280 character limit). 

Comparing the solid colored and dashed black lines across different POS tags lets us isolate the effect of the length limit on the content of the tweets. The largest gap is observed for swear words and spoken categories (nonfluencies, fillers, and assent), adverbs, and conjunctions--nonessential parts of speech that are deleted in the process of ``squeezing in'' a message to fit a length limit. No gap is observed for verbs and negations, essential parts of speech that are known to be disproportionately preserved in the shortening process ~\cite{10.1145/3359147}.

\begin{figure}[t]
    \begin{minipage}{.23\textwidth}
        \centering
        \includegraphics[width=\textwidth]{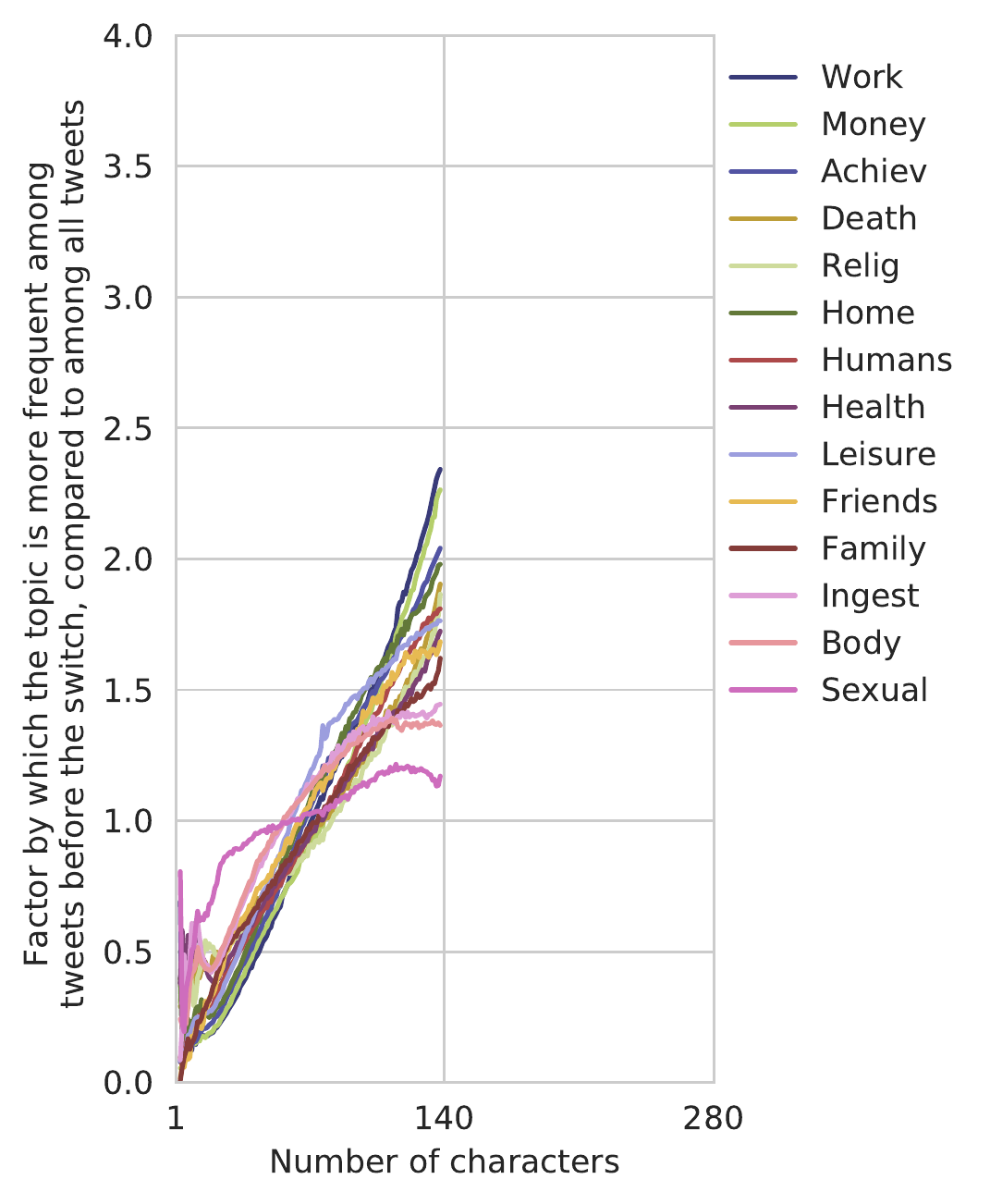}
        \label{fig:14}
    \end{minipage}
    \hfill
    \begin{minipage}{.23\textwidth}
        \centering
        \includegraphics[width=\textwidth]{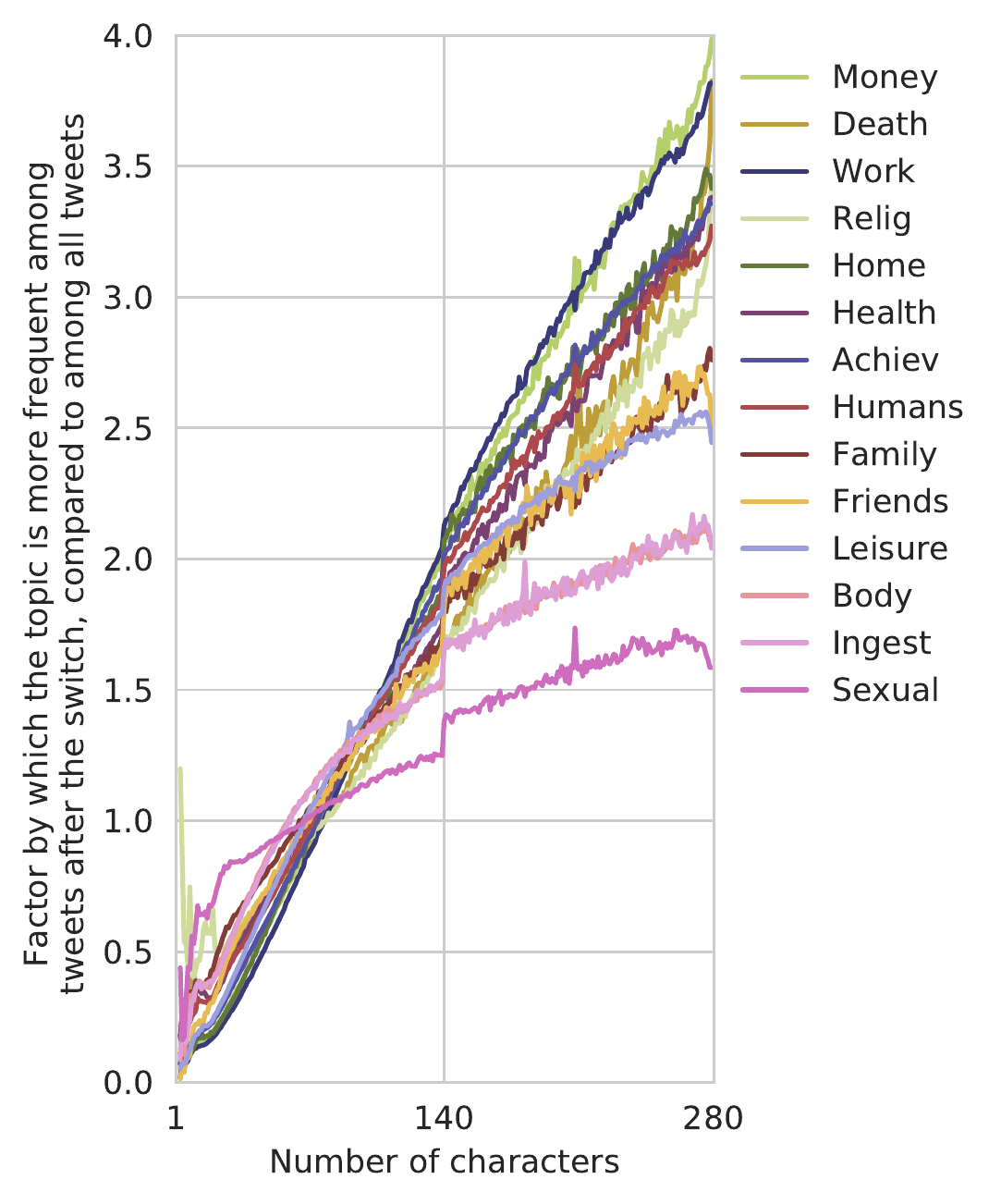}
        \label{fig:15}
    \end{minipage}
\caption{For each allowed number of characters before the switch (left), and after the switch (right), across 14 topics measured with LIWC categories, we monitor the factor by which the topic is more frequent among tweets with that number of characters, compared to overall frequency. Categories are sorted by the value of this factor at 140 characters (left), and 280 characters (right).}
\label{fig:politics}
\end{figure}

Figure~\ref{fig:11} represents the same quantities after the switch (under 280 character limit) across all possible tweet lengths in this period ($[1-280]$ characters). We note that there is no counterfactual observation, \ie, we do not know what the probability of observing a POS tag among 280 character long tweets would be if the 280 character length limit was lifted. Nonetheless, among the 280 character long tweets after the switch, we do observe patterns similar to those associated with the 140 character long tweets before the switch. We observe a ``dip'' in the frequency of the spoken categories, conjunctions, and numbers among the 280 character-long tweets, traces typical of ``optimizing'' a message to fit a length limit.

In Figures~\ref{fig:12} and~\ref{fig:13} we measure the same quantities for fine-grained subtypes of personal pronouns. This suggests that personal pronouns \textit{I} and \textit{you} were most affected by the 140 character limit (\ie, they were most likely to be omitted). We observe a similar non-monotonic distribution around 280 characters in Figure~\ref{fig:13}.

To summarize, 280 character tweets are syntactically similar to 140 tweets before the switch: their usage is associated with patterns that are indicative of ``squeezing in'' a message. This is evidence indicating that they are generated by similar writing processes as 140 character tweets were.

Next, in Figure ~\ref{fig:politics} we study topics of tweets, measured by LIWC categories describing psychological, biological, and social categories. Across the studied topics, for each allowed character length in the pre-switch and post-switch periods, we measure the factor by which the topic is more frequent in a given character length, compared to the overall topic frequency. Categories are sorted by the value of this factor at 140 characters before the switch, and at 280 characters after the switch.

We note the personal concerns categories that were relatively most frequent at 140 characters before the switch: Work, Money, Achievement, Death, and Religion. The least frequent topics at 140 characters, on the other hand, were Biological categories: Sexual, Body, Ingestion, followed by Family, Friends, and Leisure. An apparent association with the importance of the message emerges: while tweets about topics related to ordinary, overall more prevalent every-day experiences use the longer tweets the least frequently, topics related to more serious personal concerns use them the most. Similarly to within-language, there is a within-topic correlation between usage of 140 character length before the switch, and subsequent usage of 280 character length after the switch, with the ranking of the topics usage at the boundary length only slightly changed (Spearman rank correlation between topics $0.91$, $p = 7.30*10^{-6}$), implying that 280 character tweets are also semantically similar to 140 tweets before the switch.

\section{Discussion and conclusions}
\label{sec:discussion}

To summarize, immediately after the switch we observe a sharp decline in the frequency of 140 characters, and an increase in the frequency of tweets long 280 characters (Figure~\ref{fig1}). As 140 characters were becoming less prevalent after the switch, the prevalence of 280 was increasing after its introduction over a period of around 6 months (Figures~\ref{fig2} and~\ref{fig3}). Taking advantage of the emergence of differing length restriction among languages, we estimate a significant increase in tweet lengths in languages where the switch happened, compared to the control languages (Figure ~\ref{fig9}). We note, however, that tweet lengths increased slightly in the control languages as well. This is likely impacted by the nature of how tweet length is counted at the character level (as described in the Data Section), allowing mixed\hyp character tweets to be longer than 140 characters.

We observe that the more 140 was used in a language before the switch, the more 280 is used after (Figure~\ref{fig5}). Disaggregation across languages reveals interesting temporal patterns (Figure~\ref{fig7}): In most languages, the prevalence of long tweets seems to have settled, and is even decreasing again, \ie, the peak of usage, or the ``honeymoon phase'' is over. This is indicating the presence of a period of high usage rates of the new feature, followed by a drop and saturation to a constant level.

Different adoption patterns are observed between web and mobile devices (Figure~\ref{fig8}). Longer tweets are particularly used on web clients. In the web interface, 140 was quickly surpassed by 280, reaching around 4\% at the end of the studied period. While slower adoption rates on mobile devices could conceivably be linked to clients that did not update, the fact that tweets were longer on web interface before the switch indicates that there is simply a tendency for shorter text on mobile phones. Automated sources and third-party applications are the slowest to adapt. They posted longer tweets before the switch, and post tweets in longer length ranges after the switch (Figure~\ref{fig6}). It is interesting to note that automated sources and third-party applications write longer than humans, but the effect is weaker at the boundary (just under 140 before and just under 280 after the switch), probably due to the ``squeezing'' of originally longer tweets that humans do.

Tweets long 280 characters are syntactically similar to 140 tweets before the switch: their usage is associated with patterns that are indicative of ``squeezing in'' a message (Figures~~\ref{fig:10} and~\ref{fig:11}). This is evidence indicating that tweets close to the new boundary are generated by similar writing processes as 140 character tweets were before the switch. Similarly to within-language, there is a within-topic correlation between usage of 140 character length before the switch, and subsequent usage of 280 character length after the switch (Figure~\ref{fig:politics}). Tweets 280 characters long are semantically similar to 140 tweets before the switch: their usage is associated with the same topics. We note that this holds across the range of long character lengths, and not only for 280 character long tweets specifically.

Given these findings, our guiding question---\textit{Is 280 the new 140?}---calls for a nuanced answer.
On the one hand, the emergence of the peak in the prevalence of 280 characters, the fact that the languages that came close to the 140-limit also tend to come closer to the 280-limit, and the traces of distinctive writing processes when ``squeezing'' a message, absent in automated sources and third-party applications (Figure~\ref{fig6}) all resonate with a narrative stating that, yes, 280 is the new 140.
On the other hand, the prevalence of 280 is much less drastic than that of 140 used to be \cite{twitter280blog}---only 4\% for 280 characters after the switch, compared to over 12\% for 140 characters before the switch (for tweets posted by Web clients). In this sense, while 280 may indeed be considered the new 140, it is at the same time less noticeable:
in a nutshell, \textit{280 is a less intrusive 140.}



Finally, this evidence suggests that, just as the old 140\hyp character limit \cite{w1}, the new 280\hyp character limit impacts the writing style and content of tweets \cite{sen2019total}. The length constraint and the resulting tweet\hyp length distribution remain an important dimension to consider in studies using Twitter data, as after the switch the number of characters remains an important variable, correlated with important properties of tweets including topics, language, device, and the likelihood of being an automated source of tweets.

\xhdr{Limitations} This study suffers from limitations that the 1\% stream is known to be susceptible to \cite{wu2020variation}, as certain accounts might be over-represented due to the intentional or unintentional tampering with the Sample API
\cite{pfeffer2018tampering,morstatter2013sample}.

In our study, we use automated sources and third-party applications as a proxy for bots. However, bot detection can be more reliable using more sophisticated methods detecting bots that use regular applications \cite{chavoshi2016debot,kudugunta2018deep,yang2020scalable}.
Syntactic and semantic analysis of tweets is limited to tweets in English only, due to the lack of available tools to support annotation. Future studies should measure these characteristics in other languages, using language-specific tools, or machine translation.
Lastly, we note that in this study, we study Twitter as a platform (tweets are sampled at the community level), as opposed to users, whose timelines are incomplete.

\xhdr{Future work} Future work should provide a better understanding of what user features are associated with adoption (\eg, users' age, number of followers, levels of activity). Are the users who used 140 the same ones who are more likely to use 280? Answering this question requires collection of data beyond 1\% sample, that contains complete records of users' tweets. However, here we caution against naive comparisons, as careful quasi-experimental designs are necessary to truly isolate the effect of age. User age is correlated with other factors--users who stay longer on the platform might be in other ways fundamentally different from younger users who joined more recently (\ie, there is ``survivor bias'' \cite{elton1996survivor}).
Finally, our study should be replicated several years from today, when even more time has passed since the switch.



{\small
\bibliographystyle{aaai}
\bibliography{references}
}

\end{document}